\documentclass{ptptex}

\usepackage{graphicx}
\usepackage{psfrag}
\usepackage{wrapft}

\def\nn{\nonumber}
\def\be{\mbox{\boldmath $e$}}
\def\bp{\mbox{\boldmath $p$}}

\title{Local Energy Gap in Deformed Carbon Nanotubes}

\author{
Ken-ichi SASAKI,$^{1,}$\thanks{E-mail: sasaken@imr.edu}
Yoshiyuki KAWAZOE$^1$ and Riichiro SAITO$^2$ 
}

\inst{
$^1$Institute for Materials Research, Tohoku University, 
Sendai 980-8577, Japan \\
$^2$Department of Physics, Tohoku University and CREST, JST,
Sendai 980-8578, Japan 
}

\recdate{July 30, 2004}

\abst{
 The effects of graphite surface geometrical deformation on the dynamics
 of conducting electrons are investigated theoretically.
 The analysis is performed within the framework of a deformation-induced
 gauge field and corresponding deformation-induced magnetic field.
 It is shown that the latter gives a local energy gap along the axis of
 a deformed nanotube. 
 We compare our energy gap results with experimental data on energy gaps
 in nanotubes and peapods. 
 We also discuss the mixing of two Fermi points and construct a general
 model of low energy dynamics, including a short-range deformation of
 the graphite sheet.
 This model is equivalent to the Weyl equation in {\it U}(1) Abelian and
 {\it SU}(2) non-Abelian deformation-induced gauge fields.
}

\notypesetlogo
\begin{document}
\maketitle

\section{Introduction}\label{sec:intro}

The geometric structure of materials and their electronic and magnetic
properties are closely related.
It is widely recognized that carbon nanotubes~\cite{Iijima} provide us
with a great opportunity to study this relationship.
For instance, a single-wall carbon nanotube (SWNT) exhibits either
metallic or semiconducting behavior, depending on the lattice structure
around the tube axis~\cite{Wildoer}.
This structure dependent electric property can be understood by noting
that the energy band of an SWNT is determined by the wave vector
quantized by the periodic boundary condition around the tubule axis and
the energy dispersion relation of graphene~\cite{SDD}.

The energy dispersion relation of a plane of graphene is unique, because
of the linear dispersion relation at the two distinct Fermi points (the
K and K' points).
Moreover, since the unit cell of a honeycomb lattice structure consists
of two sublattices, the linear energy dispersion relation leads to
dynamics described by the Weyl equation~\cite{SW}. 

However, the dynamics of the conducting electrons of realistic
graphene-based materials are different from those of flat graphene,
because in the former case, there are shape fluctuations that result in
the modification of the overlap integral of nearest-neighbor
$\pi$-orbitals.
Even when a graphene sheet is folded into a cylinder to form a carbon
nanotube, the mean curvature of the graphite surface changes the
dynamics of the conducting electrons~\cite{Saito}.
In this case, a perturbation due to atomic deformation appears in the
dynamics through the gauge coupling~\cite{KM}. 
A conducting electron moves in the curved graphite sheet in which the
effect of curvature can be expressed in terms of a gauge field.
The effective gauge field shifts the Fermi point in the Brillouin zone
due to the Aharonov-Bohm effect and therefore changes the energy band
structure.
The effect of the finite curvature on the low-energy dynamics was
elucidated in this way by Kane and Mele~\cite{KM}.
Their theoretical prediction for the energy gap induced by the curvature
effect was confirmed by the scanning tunneling spectroscopy (STS)
experiment of Ouyang et al.~\cite{Ouyang} and was applied to different
chiral structures in several subsequent works~\cite{KE}. 

In this paper, we generalize the idea of the effective gauge field to a
local deformation of the graphite surface and show that the
deformation-induced gauge field for a locally deformed lattice provides
a local energy gap.
This gauge field is useful for describing finite-scaled nano-meter
materials.

This paper is organized as follows.
In \S\S~\ref{sec:overview} and~\ref{sec:gap-gene}, we illustrate how
the electron dynamics depend on the geometry of a deformed surface.
In particular, we consider how a local deformation affects the local
electrical properties of conducting electrons in SWNTs in terms of the
deformation-induced gauge field. 
We point out here that the deformation-induced gauge field and the usual
electro-magnetic gauge field are different, though the analogy is useful
in many cases.
Furthermore, the deformation-induced {\it magnetic} field concerns not
the spin of an electron but its pseudo-spin~\cite{ANS}, as defined by
the two sublattices of graphite and SWNTs. 
In \S~\ref{sec:basis}, we clarify the relationship between the atomic
deformation and a local energy gap structure.
We compare our theoretical result with the local energy gap observed by
STS in nanotubes~\cite{Ouyang} and peapods~\cite{Lee}.
In \S~\ref{sec:mixing}, we consider a short-range deformation that
results in a mixing of the wave functions at two Fermi points, and we
construct an effective model for the low-energy dynamics.

\section{Definition of the Gauge field for deformation}\label{sec:overview}

We consider the quantum behavior of conducting electrons on a graphite
surface using the nearest-neighbor tight-binding Hamiltonian, 
\begin{align}
 {\cal H}_{\rm near} =  \sum_a \sum_{i \in {\rm A}} 
  V_a(r_i) a_{i+a}^\dagger a_i + h.c.
\end{align}
Here, A (in the summation index) denotes an A-sublattice, $a_i$ and 
$a_i^\dagger$ are the canonical annihilation-creation operators of the 
electrons at site $i$, which satisfy the anti-commutation relation 
$\{ a_i,a_j^\dagger \} = \delta_{ij}$, and a site labeled $i+a$ (with
$a=1,2,3$) is a nearest neighbor of site $i$.
We include deformation of the graphite surface in the form of the
position-dependent hopping integral $V_a(r_i)$ in the Hamiltonian.

We decompose the hopping integral into two components as 
$V_a(r_i) \equiv V_\pi + \delta V_a(r_i)$ and define 
${\cal H}_{\rm near} = {\cal H}_0 + {\cal H}_{\rm deform}$,
where 
\begin{align}
 &
 {\cal H}_0 \equiv \sum_a \sum_{i \in {\rm A}} 
 V_\pi a_{i+a}^\dagger a_i + h.c.,
 \label{eq:hamiltonian_0} \\
 &
 {\cal H}_{\rm deform} 
 \equiv \sum_a \sum_{i \in {\rm A}} 
 \delta V_a(r_i) a_{i+a}^\dagger a_i
 + h.c.
 \label{eq:hamiltonian_1}
\end{align}
Hereafter, we refer to the {\it deformed} Hamiltonian as 
${\cal H}_{\rm deform}$. 
The total Hamiltonian adopted for this study is 
${\cal H}_{\rm near} = {\cal H}_0 + {\cal H}_{\rm deform}$, 
which is not easy to solve for general $\delta V_a(r_i)$. 
However, as long as we consider behavior near the Fermi level, we can
obtain several important physical consequences; for example, the model
Hamiltonian can exhibit a local energy gap~\cite{Lee} for a specific
deformation $\delta V_a(r_i)$.

In this paper, we mainly consider the case in which the deformation is
sufficiently delocalized in comparison with the diameter of an SWNT, so
that there is no interaction between two Fermi points.
In this case, we have doubly degenerate electronic states near the
K and K' points.
We follow the effective-mass description~\cite{LK} for low-energy
conducting electrons around each Fermi point.
Then, the Hamiltonian around the K-point, ${\cal H}_{\rm K}$, is given
by
\begin{equation}
 {\cal H}_{\rm K}
  = v_F \sigma \cdot \left( p - A \right),
  \label{eq:weyl-1}
\end{equation}
where $\sigma \cdot p = \sigma_1 p_1 + \sigma_2 p_2$
and $\sigma \cdot A = \sigma_1 A_1 + \sigma_2 A_2$.
Here, $A$ is a vector field defined on the surface, $v_F$ is the Fermi
velocity and $\sigma_i$ $(i=1,2,3)$ are the Pauli matrices, given by
\begin{align}
\sigma_1 = \begin{pmatrix} 0 & 1 \cr 1 & 0 \end{pmatrix}, \ 
\sigma_2 = \begin{pmatrix} 0 & -i \cr i & 0 \end{pmatrix}, \ 
\sigma_3 = \begin{pmatrix} 1 & 0 \cr 0 & -1 \end{pmatrix}.
\end{align}
We define coordinate axes around and along the nanotube axis as $x_1$
and $x_2$. 
The first term in Eq.~(\ref{eq:weyl-1}), $v_F \sigma \cdot p$, describes
the energy dispersion of ${\cal H}_0$~\cite{SW}, and the second term,
$-v_F \sigma \cdot A$, denotes the energy dispersion of 
${\cal H}_{\rm deform}$.
The vector field $v_F A(r)$ is a linear function of $\delta V_a(r)$.

We derive Eq.~(\ref{eq:weyl-1}) and obtain an explicit form of $A(r)$ in
the next section, but here we note two points.
First, the deformation can be included in the Hamiltonian as a gauge
field, because of the gauge coupling $p \to p-A$.
We call $A$ the {\it deformation-induced gauge field} and distinguish it
from the electro-magnetic gauge field $A^{\rm em}$.
Note that $A$ does not break the time-reversal symmetry, because
the sign in front of $A$ is different for the K and K' points.
This contrasts with the fact that $A^{\rm em}$ does violate the
time-reversal symmetry (see Eq.~\ref{eq:eff-dynamics}).
The second point is that we can decompose any gauge field into the sum
of constant, rotationless, and divergenceless components as 
\begin{align}
 A_i = A^0_i + \partial_i \Psi_a + \epsilon_{ij} \partial_j \Psi_b, 
  \ (i,j = 1,2)
  \label{eq:decomp}
\end{align}
where $\epsilon_{ij}$ is the antisymmetric tensor 
($\epsilon_{12} = - \epsilon_{21} = 1$ and $\epsilon_{ii} = 0$), and
$\Psi_a$ and $\Psi_b$ are regular scalar functions defined on the tube
surface.

Several important physical properties can be derived from 
${\cal H}_{\rm K}$ through dimensional reduction.
We integrate ${\cal H}_{\rm K}$ over the circumferential coordinate
($x_1$) to obtain a one-dimensional theory:
\begin{equation}
 \oint \frac{dx_1}{|C_h|} {\cal H}_{\rm K} = 
  v_F  (p_1^0 - A_1^0 - \delta m(x_2)) \sigma_1 + v_F (p_2 -A_2^0) \sigma_2.
  \label{eq:effective-h}
\end{equation}
Here, $|C_h|$ is the circumference of the nanotube, $p_1^0$ is
determined by the chiral index, and we have 
\begin{equation}
 \delta m(x_2) =  \oint \frac{dx_1}{|C_h|} \partial_2 \Psi_b(x_1,x_2).
  \label{eq:del-m}
\end{equation}
The function $\Psi_a$ in $A_1$ disappears in the integration over the 
$x_1$ coordinate.
We have ignored the $\Psi_a$ part of $A_2$, because it can be expressed 
as the phase of the wave function.

We identify the $\sigma_1$ term with the {\it mass} of a particle
propagating along the axis direction ($x_2$).
This is because (1) the energy spectrum at position $x_2$ is bounded by
the square root of the sum of the squares of each coefficient of the
Pauli matrices in Eq.~(\ref{eq:effective-h}) as
$\pm v_F \sqrt{ (p_1^0 - A_1^0 - \delta m(x_2))^2 + (p_2 -A_2^0)^2 }$,
and (2) $p_2$ scales as the inverse of the system length ($1/|T|$), and
hence $p_2-A_2^0$ can be ignored in comparison with 
$p_1^0 - A_1^0 -\delta m(x_2)$ for a long system ($|T| \gg |C_h|$). 
We therefore define the {\it local energy gap} as the minimum energy gap
for long tubes, given by
\begin{align}
 E_{\rm gap}(x_2) 
 \equiv 2v_F \left| p_1^0 - A_1^0 - \delta m(x_2) \right|
 = 2 v_F \left| p^0_1 -  \oint \frac{dx_1}{|C_h|} A_1(x_1,x_2) \right|.
 \label{eq:band-gap}
\end{align}
Because the circumferential component $(A_1)$ of the deformation-induced
gauge field is determined by the modulation part of the hopping
integral, knowing $\delta V_a(r)$ allows us to calculate the local
energy gap using the above formula. 
The modulation part of the hopping integral can be calculated by
choosing a specific scheme.
We adopt the Slater-Koster scheme and apply the formula to several
deformed nanotubes. 
In the next section, we give a detailed derivation and present a general
formula applicable to an arbitrary chiral index.

\section{Derivation of the local energy gap}\label{sec:gap-gene} 

There are two main purposes of this section.
One is to obtain an explicit form of $A$ in terms of $\delta V_a(r_i)$,
which is necessary to derive an explicit formula for the local energy
gap, and the other is to state a general property of the low-energy
Hamiltonian.

\begin{figure}[htbp]
 \begin{center}
  \psfrag{x}{$x$}
  \psfrag{y}{$y$}
  \psfrag{T1}{$a_1$}
  \psfrag{T2}{$a_2$}
  \psfrag{u1}{\small $R_1$}
  \psfrag{u2}{\small $R_2$}
  \psfrag{u3}{\small $R_3$}
  \includegraphics[scale=0.35]{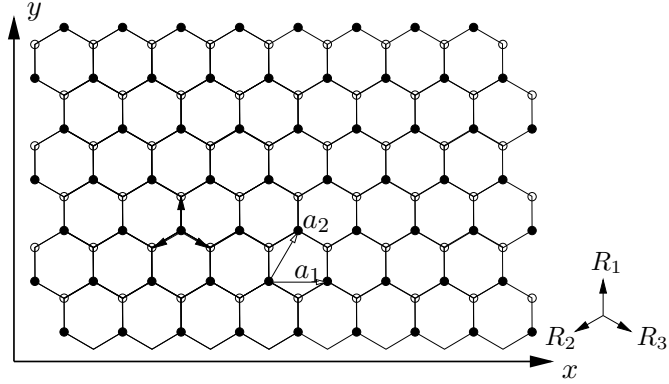}
 \end{center}
 \caption{Structure of a honeycomb lattice with the two symmetry
 translation vectors $a_1 = \sqrt{3} a_{\rm cc} e_x$ and 
 $a_2 = (\sqrt{3}/2)a_{\rm cc}e_x + (3/2)a_{\rm cc}e_y$, where 
 $e_x$ and $e_y$ are unit vectors, and $a_{\rm cc}$ is the
 nearest-neighbor bond length.  
 The black (white) circles indicate the A (B) sublattices.
 The vectors $R_a$ $(a = 1,2,3)$ point to the nearest-neighbor sites 
 of an A site.
 They are given by $R_1 = a_{\rm cc} e_y$, 
 $R_2 = -(\sqrt{3}/2) a_{\rm cc} e_x -(1/2) a_{\rm cc} e_y$ and 
 $R_3 = (\sqrt{3}/2) a_{\rm cc}e_x - (1/2) a_{\rm cc} e_y$.} 
 \label{fig:lattice}
\end{figure}

For the above-stated purposes, we first rewrite
Eq.~(\ref{eq:hamiltonian_0}) using the Bloch basis vectors as
\begin{align}
 {\cal H}_0 = V_\pi \sum_k f(k)
 |\Psi_A^k \rangle \langle \Psi_B^k | + h.c.,
 \label{eq:near}
\end{align}
where we define $f(k) \equiv \sum_a f_a(k)$ and 
$f_a(k) \equiv e^{ik \cdot R_a}$.
Here, $R_a$ $(a = 1,2,3)$ are vectors pointing to the nearest-neighbor
sites from an A site (Fig.~\ref{fig:lattice}).
We have defined the Bloch basis vectors with wave vector $k$ as 
\begin{align}
 |\Psi_\alpha^k \rangle = \frac{1}{\sqrt{N_\alpha}} 
 \sum_{i \in \alpha} e^{ik \cdot r_i} 
 a_i^\dagger |0 \rangle, 
 \label{eq:bloch}
\end{align}
where the subscript $\alpha$ ($=A,B$) denotes two sublattices, as
illustrated in Fig.~\ref{fig:lattice}, and $N_A$ ($=N_B$) denotes the
number of black (white) sites represented by solid (empty) circles.
Here, $r_i$ labels the vector pointing toward the site $i$.

We expand $f(k)$ in Eq.~(\ref{eq:near}) around the K point, $k_F$, as
\begin{align}
 f(k) = 
 \sum_a f_a(k_F) i(k-k_F) \cdot R_a 
 + \cdots,
 \label{eq:off-diag-low}
\end{align}
where we use the condition for the Fermi point $f(k_F)= 0$. 
[We can set $f_1(k_F) = 1$, $f_2(k_F) = e^{-i \frac{2\pi}{3}}$ and  
$f_3(k_F) = e^{+i \frac{2\pi}{3}}$.]
The lattice structure of SWNTs is specified by the chiral and
translational vectors defined by 
$C_h = n a_1 + m a_2$ and $T = p a_1 + q a_2$, 
where $a_1$ and $a_2$ are translational symmetry vectors in the planar
honeycomb lattice (Fig.~\ref{fig:lattice}).
The corresponding wavevector $k$ can be decomposed as 
$k= \mu_1 k_1 + \mu_2 k_2$, where $\mu_1$ and $\mu_2$ are integer
coefficients of the vectors $k_1$ and $k_2$, which satisfy 
$C_h \cdot k_1 = 2\pi$, $C_h \cdot k_2 = 0$, $T \cdot k_1 = 0$ and 
$T \cdot k_2 = 2\pi$. 
With these definitions, the first term on the right-hand side of
Eq.~(\ref{eq:near}) is approximated by
\begin{align}
 V_\pi e^{-i\theta} w|\Psi_A^k \rangle \langle \Psi_B^k|,
  \label{eq:eff-1}
\end{align}
where $w \equiv w_1 -iw_2$, $w_1$ and $w_2$ are given by
\begin{align}
 w_1 = \frac{3a_{\rm cc}}{2|C_h|}(2\pi \mu_1 - k_F \cdot C_h), \ \
 w_2 = \frac{3a_{\rm cc}}{2|T|}(2\pi \mu_2 - k_F \cdot T),
\end{align}
and the angle $\theta$ satisfies
\begin{align}
 \left(n+\frac{m}{2}\right) - i \frac{\sqrt{3}}{2} m  = 
  \frac{|C_h|}{\sqrt{3}a_{\rm cc}}e^{-i\theta}.
\end{align}
To obtain Eq.~(\ref{eq:eff-1}), it is useful to rewrite $R_a$ in terms
of the chiral and translational vectors as
\begin{align}
 R_1 &= \frac{2}{3N_c} 
  \left[ \left(p+\frac{q}{2}\right) C_h - \left(n+\frac{m}{2}\right) 
   T \right], \nn \\
 R_2 &= \frac{2}{3N_c} 
  \left[ -\left(\frac{p-q}{2}\right) C_h + \left(\frac{n-m}{2}\right)
   T \right], \nn \\
 R_3 &= \frac{2}{3N_c} 
  \left[ -\left(\frac{p}{2} + q \right)C_h + \left(\frac{n}{2} + m\right) 
   T \right], \label{eq:R_a}
\end{align}
where we introduce $N_c \equiv mp-nq$ and use the relationship
$T \times C_h = (3\sqrt{3}/2)a_{\rm cc}^2 N_c (e_x \times e_y)$.  
Finally, the leading term in the Hamiltonian matrix for ${\cal H}_0$ can
be expressed in the basis of the two sublattices ${}^t(\Psi_A,\Psi_B)$
as 
\begin{align}
 {\cal H}_0 
  = V_\pi 
  \begin{pmatrix} 
  0 & e^{-i\theta} w \cr e^{i\theta} w^* & 0 
  \end{pmatrix}.
  \label{eq:eff-hami}
\end{align}

Next, we consider the momentum transfer matrix element of 
${\cal H}_{\rm deform}$,
\begin{align}
 \langle \Psi^{k+\delta k}_A | {\cal H}_{\rm deform} | \Psi^k_B
 \rangle = \frac{1}{N_A} \sum_{i \in A} \sum_a
 \delta V_a(r_i) f_a(k) e^{-i \delta k \cdot r_i},
\end{align}
where we have used Eqs.~(\ref{eq:hamiltonian_1}) and (\ref{eq:bloch}) to
obtain the right-hand side.
Here, we restrict ourselves to considering the momentum transfer matrix
element that does not mix the two Fermi points 
(i.e. we assume $|\delta k| \ll |k_F|$).
We then set $k = k_F + \delta k'$ and obtain
\begin{align}
 \langle \Psi^{k+\delta k}_A | {\cal H}_{\rm deform} | \Psi^{k}_B
 \rangle 
 = \frac{1}{N_A} \sum_{i \in A} \left( \sum_a \delta V_a(r_i) f_a(k_F)
 \right) e^{-i \delta k \cdot r_i},
 \label{eq:defo}
\end{align}
where we omit an ${\cal O}\left( \delta V_a \delta k' \right)$
correction.
The momenta $k$ and $k + \delta k$ in the last equation should be
recognized as being restricted to the wave vector around the K point.
Up to leading order, this shows that the electrons near the K point
are subject to the ``local potential'' given by 
$\sum_a \delta V_a(r_i) f_a(k_F)$, because its Fourier component appears
in the right-hand side of Eq.~(\ref{eq:defo}). 
It is important to note that there is a similarity between the
deformation Hamiltonian and the potential $V(x)$ (whose constant
component should be subtracted) for a particle of mass $m$. 
Let us consider a simple one-dimensional Hamiltonian,
${\cal H} = \frac{p^2}{2m} + V$.
When we consider the matrix element of the potential term between free
particle states denoted $|k \rangle$ with a definite momentum, we obtain
$\langle k+\delta k | V | k \rangle = \int \frac{dx}{L} V(x) 
e^{-i \delta k \cdot x}$,
where $L$ is the system length.
This should be compared with Eq.~(\ref{eq:defo}), and it shows that 
$\sum_a \delta V_a(r_i) f_a(k_F)$ acts as a potential.
Note also that Eq.~(\ref{eq:eff-hami}) corresponds to the kinetic term
of Eq.~(\ref{eq:near}).

Combining Eq.~(\ref{eq:eff-hami}) with Eq.~(\ref{eq:defo}), we obtain
the effective-mass Hamiltonian
\begin{align}
 & V_\pi 
 \begin{pmatrix} 0 & e^{-i\theta} w \cr e^{i\theta} w^* & 0 \end{pmatrix}
 - v_F \sigma \cdot A  \nn \\
 &=
 \begin{pmatrix} e^{-i\theta} & 0 \cr 0 & e^{i\theta} \end{pmatrix}
 v_F \left[  \sigma \cdot p - 
 \begin{pmatrix} e^{+i\theta} & 0 \cr 0 & e^{-i\theta} \end{pmatrix}
 \sigma \cdot A \right], 
 \label{eq:eff-2}
\end{align}
where we have denoted $\sum_a \delta V_a(r_i) f_a(k_F)$ as 
$-v_F (A_x-iA_y)$ and used $f_1(k_F) = 1$, 
$f_2(k_F) = e^{-i \frac{2\pi}{3}}$
and $f_3(k_F) = e^{+i \frac{2\pi}{3}}$.
Here, we have defined the Fermi velocity as 
$v_F = 3V_\pi a_{\rm cc}/2\hbar$, and we have
$\sigma \cdot p = (2\hbar/3a_{\rm cc}) (\sigma_1 w_1 + \sigma_2 w_2)$
and $\sigma \cdot A = \sigma_1 A_x + \sigma_2 A_y$.
Then, functions $A_x$ and $A_y$ are given by
\begin{align}
 - v_F A_x = 
 \delta V_1 - \frac{1}{2} \delta V_2 - \frac{1}{2} \delta V_3,  \ 
 - v_F A_y = \frac{\sqrt{3}}{2} \left( \delta V_2 - \delta V_3 \right).
 \label{eq:p-gauge-A}
\end{align}
This expresses the explicit relationship between the hopping integral
modulation and the deformation-induced gauge field. 
We refer to functions $A_x$ and $A_y$, or their appropriate linear
combination, as the deformation-induced gauge field.
This is because the effects of the deformation can be included in the
theory through the substitution $p \to p -A$. 
This is the same substitution as for an electro-magnetic gauge field.

We simplify the notation in Eq.~(\ref{eq:eff-2}) and state a general
property of the effective Hamiltonian.
We have two distinct modes, whose dynamics are approximated by 
the following two effective Hamiltonians in the presence of an 
{\it external} electro-magnetic gauge field $A^{\rm em}$:
\begin{align}
 {\cal H}_{\rm K}
 = v_F \sigma \cdot \left( p - A - A^{\rm em} \right),\ \  
 {\cal H}_{\rm K'}
 = v_F \sigma' \cdot \left( p + A - A^{\rm em} \right).
 \label{eq:eff-dynamics}
\end{align}
Here, we have defined $\sigma = (\sigma_1,\sigma_2)$ and 
$\sigma' = (-\sigma_1,\sigma_2)$.
The Hamiltonian ${\cal H}_{\rm K'}$ can be obtained by expanding
Eq.~(\ref{eq:near}) around the K' point, $-k_F$, and repeating the
calculation presented above. 
Our notation for the momentum is 
$\sigma \cdot p = \sigma_1 p_1 + \sigma_2 p_2$ and 
$\sigma' \cdot p = -\sigma_1 p_1 + \sigma_2 p_2$, where
\begin{align}
 p_1 =  \frac{\hbar (2\pi \mu_1 - k_F \cdot C_h)}{|C_h|}, \ \
 p_2 =   \frac{\hbar(2\pi \mu_2 - k_F \cdot T)}{|T|}, 
 \label{eq:momentum} 
\end{align}
and our notation of the deformation-induced gauge field is
$\sigma \cdot A = \sigma_1 A_1 + \sigma_2 A_2$ and 
$\sigma' \cdot A = -\sigma_1 A_1 + \sigma_2 A_2$, where
\begin{align}
 \begin{pmatrix} A_1 \cr A_2 \end{pmatrix} = 
 \begin{pmatrix} 
 \cos \theta & \sin \theta \cr - \sin \theta & \cos \theta 
 \end{pmatrix}
 \begin{pmatrix} A_x \cr A_y \end{pmatrix}.
 \label{eq:A1}
\end{align}
Note that $p_1$ and $p_2$ are, respectively, the momenta around and
along the axis as measured from the K point. 
For the K' point, $k_F$ and $\mu_1$ should change sign, and,
correspondingly, $p_1$ for the K' point will have the sign opposite to
that of the K point.
Note also that the sign of $A$ is different for the K and K' points, in
contrast to the case for $A^{\rm em}$ in Eq.~(\ref{eq:eff-dynamics}). 
Therefore, in the absence of an external electro-magnetic gauge field,
time-reversal symmetry is preserved even in the presence of a
deformation.

By virtue of the decomposition given in Eq.~(\ref{eq:decomp}) with the
gauge coupling structure of Eq.~(\ref{eq:eff-dynamics}), the following
two points become clear.
First, the $\Psi_a$ component does not change the energy spectra of the
theory, because it can be eliminated by multiplying the wave function by
a phase (i.e., taking $\psi \to e^{i\frac{\Psi_a}{\hbar}} \psi$). 
However, it should be mentioned that a nontrivial functional form of
$\Psi_a$ gives a nonvanishing divergence of the deformation-induced
gauge field, represented by $\nabla \cdot A = \nabla^2 \Psi_a \ne 0$.
This may correspond to the deformation potential and could cause a 
local (background) charge modulation~\cite{SA}.
The effective one-dimensional quantum field theory~\cite{Tomonaga}
derived from Eq.~(\ref{eq:eff-dynamics}) can be used to study this
effect.
Analysis using this theory shows that the $\Psi_a$ component does not
change the energy spectrum of the theory but gives only a charge density
modulation for metallic energy band structure cases.
Second, the function $\Psi_b$ operates as a source of a
deformation-induced {\it magnetic} field and can lead to an important
physical effect.
The curl part of the deformation-induced gauge field defines a
deformation-induced magnetic field as
\begin{equation}
 B_\perp = \epsilon_{ij} \partial_i A_j = \nabla^2 \Psi_b,
  \label{eq:mag-field}
\end{equation}
whose direction is perpendicular to the graphite surface, and a
non-trivial magnetic field gives a local modulation of the energy gap,
as defined in Eqs.~(\ref{eq:del-m}) and (\ref{eq:band-gap}).

In the previous section, in Eq.~(\ref{eq:effective-h}) we integrated
over the circumferential coordinate ($x_1$) of
Eq.~(\ref{eq:eff-dynamics}).
In Eq.~(\ref{eq:effective-h}), $p_1^0$ is determined by setting 
$\mu_1 = \mu_1^0$ in Eq.~(\ref{eq:momentum}), and $\mu_1^0$ is related
to the chiral index as $\mu_1^0 = \langle (2n+m)/3 \rangle$ 
(where $\langle x \rangle$ represents the closest integer to the real
number $x$).
Note that we have already ignored the term
$\partial_2 \oint \frac{dx_1}{|C_h|} \Psi_a(x_1,x_2)$
in Eq.~(\ref{eq:effective-h}) by selecting a proper phase for the wave
function.
The coefficient of $\sigma_1$ plays the role of the effective mass of a
particle propagating along the axis. 
Therefore, $p_1^0 - A_1^0$ is related to a position-independent mass,
and $\delta m(x_2)$ represents a local modulation of the mass. 
They combine to give the local energy gap given in
Eq.~(\ref{eq:band-gap}).

\begin{figure}[htbp]
 \begin{center}
  \psfrag{a}{(a) Energy gap}
  \psfrag{b}{(b) Deformation-induced magnetic field}
  \psfrag{x}{$x_2$}
  \psfrag{u}{$+v_F \left| p_1^0 - A_1^0 - \delta m(x_2) \right|$}
  \psfrag{d}{$-v_F \left| p_1^0 - A_1^0 - \delta m(x_2) \right|$}
  \psfrag{g}{$E_{\rm gap}(x_2)$}
  \psfrag{p}{$B_\perp$}
  \psfrag{q}{$B_\parallel$}
  \includegraphics[scale=0.35]{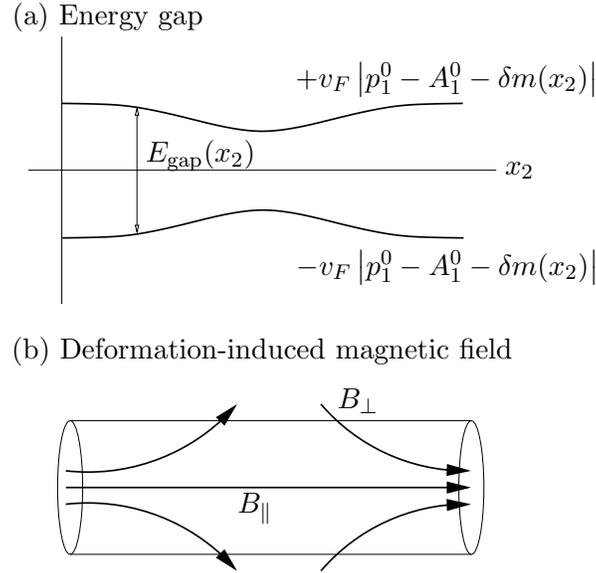}
 \end{center}
 \caption{The relationship between (a) the local energy gap and (b) the
 corresponding deformation-induced magnetic fields $B_\parallel$ and
 $B_\perp$. 
 It is noted that this is not the usual electro-magnetic magnetic field.} 
 \label{fig:field}
\end{figure}

It is important to note that the position-dependent mass (or local
energy gap) corresponds to the existence of a local deformation-induced
{\it magnetic} field, because if a nonvanishing local energy gap
remains, then its first derivative in the axis direction ($x_2$) is also
a nonvanishing value.
This corresponds to the magnetic field integrated along the $x_1$
direction as 
\begin{equation}
  \partial_2 \delta m(x_2) = \oint \frac{dx_1}{|C_h|} \partial_2^2 \Psi_b 
   = \oint \frac{dx_1}{|C_h|} B_\perp(x_1,x_2).
\end{equation}
The Stokes theorem allows us to imagine a deformation-induced magnetic
field $B_\parallel$ in a cylinder, defined by the loop integral of the
deformation-induced gauge field around the axis, given by 
\begin{equation}
 B_\parallel(x_2) S = \oint dx_1 A_1(x_1,x_2),
\end{equation}
where $S$ denotes the cross-sectional area of the cylinder, assumed to
be constant along the axis.
When the magnetic field escapes from within the cylinder, it penetrates
the surface. 
This corresponds to $B_\perp$ on the surface (penetrating the surface),
and therefore it changes the mass of the particle (see
Fig.~\ref{fig:field}). 

We now briefly consider the interesting case in which 
$p_1^0 - A_1^0 - m(x_2) - A_1^{\rm em}$ changes sign at some position.
It is known that an unusual solution, the Jackiw-Rebbi modes~\cite{JR}
(electron-deformation bound states), exists at the band center. 
It is likely that such a condition will be satisfied for the bulk part
of some types of peapod carbon nanotubes.

\section{Examples of Deformation-Induced Gauge Fields}\label{sec:basis}

In this section, we apply the theory to a curved surface of a SWNT.
First, we classify deformations into two types, using the Slater-Koster
scheme. 
The Slater-Koster scheme defines the hopping integral between
nearest-neighbor sites as
$V_a(r_i) = V_\pi(a_{\rm cc} + \delta r_a) \be(r_{i+a}) \cdot \be(r_i)$,
which can include the fact that conducting electrons form a
$\pi$-orbital whose wave function extends in the normal direction of the
surface.
Here, $V_\pi(r)$ is a function of the bond length $r$ 
[$V_\pi(a_{\rm cc}) = V_\pi$] and $\be(r_i)$ is a unit normal vector at
$r_i$. 
We have assumed here that the effect of the $\sigma$-bond is included
in the definition of $V_\pi(r)$.

\begin{figure}[htbp]
 \begin{center}
  \psfrag{a}{$\be(r_i)$}
  \psfrag{b}{$\be(r_i+R_a)$}
  \psfrag{c}{Independent on $i$}
  \psfrag{D}{$a_{\rm cc}+{\rm Eq.(\ref{eq:class-1})}$}
  \psfrag{d}{$a_{\rm cc}+\delta r_a(r_i)$}
  \psfrag{e}{Surface}
  \psfrag{B}{(a)}
  \psfrag{C}{(b)}
  \psfrag{i}{$i$}
  \psfrag{j}{$i+a$}
  \psfrag{g}{Dependent on $i$}
  \includegraphics[scale=0.25]{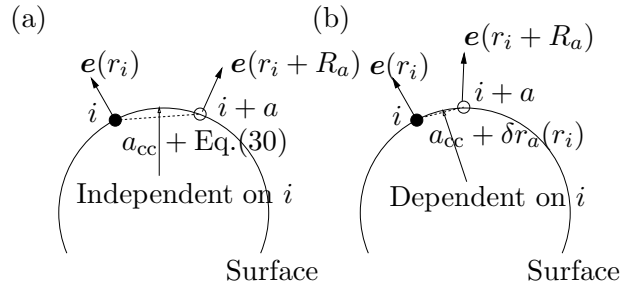}
 \end{center}
 \caption{A schematic diagram of the deformed surface of a
 two-dimensional graphite sheet.
 In these diagrams, only two sites ($\bullet$ and $\circ$) are depicted.
 We denote the direct distance between them by $a_{\rm cc} + \delta r_a$.
 We refer to (a) as the {\it bond-bending} type and (b) as the 
 {\it bond-stretching} type in this paper. 
 In (a), the direct distance is given by $a_{\rm cc}$ +
 Eq.~(\ref{eq:class-1}), and in (b) it is not related to the normal
 vector.} 
 \label{fig:bond}
\end{figure}

Now, $\be(r_i + R_a)$ can be expanded in a Taylor series around 
$\be(r_i)$ as 
\begin{align}
 \be(r_i+R_a) 
 = \be(r_i) + (R_a \cdot \nabla) \be(r_i)
 + \frac{1}{2} (R_a \cdot \nabla)(R_a \cdot \nabla) \be(r_i) + \cdots.
\end{align}
Using the normalization condition of the normal vector $\be(r_i) \cdot
\be(r_i) = 1$, we obtain 
$\be(r_i) \cdot ( R_a \cdot \nabla) \be(r_i) = 0$. 
We can then write $\delta V_a(r_i) \equiv V_a(r_i) -V_\pi$ as 
\begin{align}
 \delta V_a(r_i) 
 = \delta r_a(r_i) (\partial V_\pi|_{a_{\rm cc}})
 + V_\pi \frac{1}{2} \be(r_i) (R_a \cdot \nabla)(R_a \cdot \nabla)
 \be(r_i) + \cdots,
 \label{eq:hopp-int}
\end{align}
where $\cdots$ represents higher-order corrections.
For nanotubes, the derivative ($\nabla$) is scaled by the inverse of
the circumference ($|C_h|$) or by the inverse of the length ($|T|$),
depending on the direction.  
We can therefore consider, as an order-of-magnitude estimation, the
relevant energy of the second term on the right-hand side of
Eq.~(\ref{eq:hopp-int}) to be proportional to the inverse square of
the circumference.
(More specifically, we approximate it as the value 
$V_\pi a_{\rm cc}^2/|C_h|^2$.) 
This contributes a small (but observable~\cite{Ouyang}) correction to
the energy gap.
However, the bond length modulation, denoted by $\delta r_a(r_i)$, may
give a dominant contribution to the energy gap~\cite{Lee}. 

Here, we define two types of deformation depending on the functional
form of $\delta r_a(r_i)$.  
The first type of geometry includes the deformed surfaces of SWNTs in
which the bond length modulation between adjacent carbon atoms, as shown
in Fig.~\ref{fig:bond}(a), is defined by 
\begin{align}
 \delta r_a = \frac{a_{\rm cc}}{12}
 \left[ \frac{1}{2} 
 \be(r_i) (R_a \cdot \nabla)(R_a \cdot \nabla) \be(r_i)
 \right] + \cdots.
 \label{eq:class-1}
\end{align}
By substituting this into Eq.~(\ref{eq:hopp-int}), we obtain 
\begin{align}
 \delta V_a(r_i) = D_\pi 
 (R_a \cdot \nabla) \be(r_i) \cdot (R_a \cdot \nabla) \be(r_i), 
 \label{eq:p-gauge}
\end{align}
where we have introduced the constant
\begin{align}
 D_\pi = - \frac{1}{2} 
 \left[
 \frac{1}{12}a_{\rm cc} \partial V_\pi|_{a_{\rm cc}} + V_\pi
 \right]
 \label{eq:D-pi}
\end{align}
for the sake of convenience.
The numerical value of $D_\pi$ can be estimated from the experimental
data of Ouyang et al.~\cite{Ouyang} as $|D_\pi| = V_\pi/8$, with
$V_\pi = 2.60$ eV.
We refer to deformed surfaces whose hopping integral is approximated by 
Eq.~(\ref{eq:p-gauge}) with an appropriate $D_\pi$ as the 
{\it bond-bending} deformation type. 
The hopping integral of the bond-bending type is calculated from the
normal vectors on the surface, which means that we only need to
parameterize the surface and do not need to perform microscopic
calculations to obtain the hopping integral.
Note we assume here that the (indirect) distance between two adjacent
atoms defined by the line integral of the differential line element on a
deformed graphite surface is fixed by $a_{cc}$ and is independent of
its position.

In another type of deformation, the dominant contribution to $\delta
V_a(r_i)$ comes from a change in the bond length. 
For this type, the hopping integral is approximated by the first
term on the right-hand side of Eq.~(\ref{eq:hopp-int}) as 
\begin{align}
 \delta V_a(r_i) = 
 (\partial V_\pi|_{a_{\rm cc}}) \delta r_a(r_i) + \cdots.
 \label{eq:class-2}
\end{align}
Equation~(\ref{eq:class-2}) may be relevant to the conducting
electrons in a peapod-like structure~\cite{Lee}, in which the elastic
strain is believed to modify the bond length.
We refer to deformed surfaces whose hopping integral is approximated by 
Eq.~(\ref{eq:class-2}) as the {\it bond-stretching} deformation type.
The bond length (and its hopping integral) can be calculated with a
first-principle calculation.

\begin{table}
 \caption{Deformation types and expected energy scales. 
 The bond-bending type can be applied to ordinary single-wall nanotubes,
 and the bond-stretching type is believed to be effective for peapod 
 nanotubes.}
 \begin{center}
  \def\arraystretch{1.5}
 \begin{tabular}{|l|l|l|l|} \hline
  {\bf Type} & $\delta V_a(r_i)$ & {\bf Target} & {\bf Energy scale} \\ \hline
  Bond-bending & Eq.~(\ref{eq:p-gauge}) &
  Nanotubes & $D_\pi a_{\rm cc}^2/|C_h|^2$ \\  \hline
  Bond-stretching & Eq.~(\ref{eq:class-2}) & Peapod, etc. & 
  $(a_{\rm cc} \partial V_\pi|_{a_{\rm cc}}) \delta r_a/a_{\rm cc}$ \\ \hline
 \end{tabular}
 \end{center}
 \label{table:list}
\end{table}
Table~\ref{table:list} lists the deformation types.
A typical energy of $D_\pi a_{\rm cc}^2/|C_h|^2$ is expected for a
deformed Hamiltonian of the bond-bending type, and an energy of 
$(a_{\rm cc} \partial V_\pi|_{a_{\rm cc}}) \delta r_a/a_{\rm cc}$ for
the bond-stretching type.

\subsection{Application to several geometries}\label{subsec:application}

In this subsection, we present examples of the deformation-induced gauge
field with calculated results for the local energy gap.
We calculate the deformation-induced gauge field for several geometries,
nanotube and peapod, each of which is classified as a bond-bending or
bond-stretching deformation. 
We then apply the deformation to the local energy gap formula and compare
the results with experimental data.
We also briefly discuss the effects of the on-site energy.

\subsubsection{Narrow energy gap in a metallic zigzag nanotube
   (bond-bending type)}\label{subsec:b-class}

For geometries with a bond-bending deformation, we can apply
Eq.~(\ref{eq:p-gauge}) to the deformation-induced gauge field. 
In this case, by substituting Eq.~(\ref{eq:R_a}) into
Eq.~(\ref{eq:p-gauge}), and using Eq.~(\ref{eq:p-gauge-A}), we obtain 
\begin{align}
 v_F A_x &= F_x^{uu} \be_u \cdot \be_u + F_x^{uv} \be_u \cdot \be_v 
  + F_x^{vv} \be_v \cdot \be_v, \\
 v_F A_y &= F_y^{uu} \be_u \cdot \be_u + F_y^{uv} \be_u \cdot \be_v 
  + F_y^{vv} \be_v \cdot \be_v,
\end{align}
where we have introduced the following quantities:
\begin{align}
 & F_x^{uu} \equiv F \left( p^2+pq-\frac{1}{2}q^2 \right), \ \ \ \ 
 F_y^{uu} \equiv -F \sqrt{3}q \left( p+\frac{1}{2}q \right), \nn \\
 & F_x^{uv} \equiv -F (2pn + qn + pm -qm), \ \ 
 F_y^{uv} \equiv F \sqrt{3}(qn + pm + qm), \nn \\
 & F_x^{vv} \equiv= F \left( n^2+nm-\frac{1}{2}m^2 \right), \ \ \ \ 
 F_y^{vv} \equiv - F \sqrt{3} m \left( n+\frac{1}{2}m \right). 
 \label{eq:F_xy}
\end{align}
In the above equations, we define $F \equiv 4\pi^2 D_\pi/3N_c^2$ and
the dimensionless coordinates $(u,v)$ through the relations
$C_h \cdot \nabla = |C_h| \partial_1 = 2\pi \partial_u$ and 
$T \cdot \nabla = |T| \partial_2 = 2\pi \partial_v$.
Finally, $\be_u$ and $\be_v$ denote $\partial_u \be$ and 
$\partial_v \be$.
The most dominant contributions to the deformation-induced gauge field
are contributed by the $F_x^{uu}$ and $F_y^{uu}$ terms. 
This is because $F$ itself scales as the inverse square of the nanotube
surface area, $p$ and $q$ scale as $|T|$, and $n$ and $m$ scale as
$|C_h|$. 
Therefore, each component scales as
\begin{align}
 F_i^{uu}:F_i^{uv}:F_i^{vv}=
  \frac{a_{\rm cc}^2}{|C_h|^2}:
  \frac{a_{\rm cc}^2}{|C_h||T|}:
  \frac{a_{\rm cc}^2}{|T|^2}, \ \ 
  i =x,y.
\end{align}
This scaling indicates that we can ignore the $F_i^{uv}$ and $F_i^{vv}$ 
terms if $|T| \gg |C_h|$. 

Because the chiral and translational indices determine each component
listed in Eq.~(\ref{eq:F_xy}), specifying the normal vector $\be$ on a
specific surface fixes the deformation-induced gauge field completely.
We can specify the surface of a tube with a constant radius by the
vector
$\bp(u,v) = \left( |C_h|/2\pi \cos u , |C_h|/2\pi \sin u , v \right)$.
The unit normal vector of this surface is calculated as
$\be = ( \cos u , \sin u, 0)$.
We then obtain 
$\be_u \cdot \be_u = 1,\be_u \cdot \be_v = 0,\be_v \cdot \be_v = 0$,
yielding the deformation-induced gauge fields
$v_F A_x = F^{uu}_x$ and $v_F A_y = F^{uu}_y$.
To calculate each component of Eq.~(\ref{eq:F_xy}) explicitly,
we use the example of zigzag nanotubes, whose chiral index is
$(n,0)$, with {\it metallic} index $p_1^0 = 0$ 
(where $n$ is a multiple of 3). 
In this case, Eq.~(\ref{eq:F_xy}) gives
$F^{uu}_x = - \pi^2 D_\pi/n^2$ and $F^{uu}_y = 0$.
Applying this result to the formula for the energy gap of a {\it
metallic} zigzag nanotube, we obtain $E_{\rm gap}(v) = 2|F^{uu}_x|$.
We now compare this with the experimental result of Ouyang et
al.~\cite{Ouyang}.
The experiment shows that the energy gap for {\it metallic} zigzag SWNTs
with $n=9$,$12$,$15$ can be fitted by 
$E_{\rm gap}^{\rm exp} = \pi^2 V_\pi/4n^2$,
where $V_\pi = 2.60$ eV,
which corresponds to $|D_\pi| = V_\pi/8$.
Using the explicit form of the parameter $D_\pi$, we can read off the
first derivative of $V_\pi(r)$ as
\begin{equation}
 a_{\rm cc} \partial V_\pi|_{a_{\rm cc}} =
  \begin{cases}
   -9V_\pi \ \ \  {\rm if} \ \ D_\pi < 0, \\
   -15V_\pi \ \   {\rm if} \ \ D_\pi > 0.
   \end{cases}
  \label{eq:bond-scaling}
\end{equation}
This can be used to analyze the other chiral structures and also for
geometries of the bond-stretching deformation.
It is noted that these values differ significantly from that estimated
with the scheme employed in Refs.~\citen{Xu} and~\citen{Jiang},
$a_{\rm cc} \partial V_\pi|_{a_{cc}} \sim -3 V_\pi$.
Elucidating the reason for this discrepancy~\cite{SA} is beyond the
scope of our low energy theory.
However, we point out that the overall factor of $E_{\rm gap}^{\rm exp}$
may be affected by the details of the STS experiments. 
The important point is that the $n$ dependence of the energy gap is the
same for the experiment (although there exist data for only three
different values of $n$) and theory. 
Another possible reason for this discrepancy is that actual nanotubes
are actually between bond-bending and bond-stretching classes.
We can introduce into Eq.~(\ref{eq:D-pi}) a phenomenological parameter
$\alpha$ to describe such a nanotube as
$D^\alpha_\pi = - \frac{1}{2} \left[ \frac{\alpha}{12}a_{\rm cc}
\partial V_\pi|_{a_{\rm cc}} + V_\pi \right]$.
Then, we can choose $\alpha \sim 3$ to get 
$(a_{\rm cc} \partial V_\pi|_{a_{cc}}) \sim -3 V_\pi$.

\subsubsection{Peapod (bond-stretching type)}\label{subsec:s-class}

We consider the peapod geometry as an example of a bond-stretching
deformation whose hopping integral depends strongly on the bond length
modulation.  
As with some (encapsulated metal fullerene) peapod-like
structures~\cite{Lee}, elastic strain is expected to modify the bond 
length.
We do not attempt to calculate the bond length for such systems using
a first-principle calculation, but instead we wish to estimate the bond
length modulation (necessary for explaining an observed local energy
gap modulation within our theoretical framework).

We consider zigzag nanotubes $(n,0)$ of constant radius and assume
that $A_1 = A_x$ depends only on the axis coordinate. 
In this case, we can simplify the formula for the energy gap into the
form $E_{\rm gap}(y) = 2  v_F \left|p^0_x - A_x(y) \right|$.
From Eqs.~(\ref{eq:p-gauge-A}) and (\ref{eq:class-2}), the
deformation-induced gauge field around the axis can be written
\begin{align}
 - v_F A_x(y) =
 (\partial V_\pi|_{a_{cc}}) 
 \left( \delta r_1(y) - \frac{1}{2} \delta r_2(y)
 - \frac{1}{2} \delta r_3(y) \right).
\end{align}
It is reasonable to suppose that the bonds, pointing in three different
directions, are related, because the modulation generates a force
between adjacent atoms.  
For zigzag nanotubes, we assume the relation
$\delta r_2(y) = \delta r_3(y) = \frac{1}{2} \delta r_1(y)$, 
which gives an equilibrium configuration and preserves the rotational
symmetry about the axis.
In this case, the energy gap formula becomes
\begin{align}
 E_{\rm gap}(y) = 
 2 \left| v_F p^0_x - \frac{1}{2}(a_{\rm cc} \partial V_\pi|_{a_{cc}}) 
 \frac{\delta r_1(y)}{a_{\rm cc}} \right|,
\end{align}
where the first term on the right-hand side is given for zigzag
nanotubes with chiral index $(n,0)$ by
\begin{align}
 v_F p_x^0 = V_\pi \frac{2\pi \sqrt{3}}{n} 
 \left( \left\langle \frac{2n}{3} \right\rangle - \frac{2n}{3} \right).
\end{align}
To realize an observed energy gap modulation on the order of 0.4
eV~\cite{Lee}, the maximum bond-length modulation should be on the order
of $\delta r_1^{\rm max} \sim a_{\rm cc}/50$, where we have used the
first case in Eq.~(\ref{eq:bond-scaling}).
Although this quantity was fixed by the experimental data of Ouyang et
al.~\cite{Ouyang}, it can be estimated using another scheme~\cite{Xu} as
$a_{\rm cc} \partial V_\pi|_{a_{cc}} \sim -3 V_\pi$, in which case
the maximum bond length modulation should be about $a_{\rm cc}/20$. 
In any case, the small bond length modulation gives a rather strong
contribution to the local energy gap. 
Here, we should note that Cho et al.~\cite{Cho} accounted for the local
energy gap observed by Lee et al.~\cite{Lee} in terms of the gap between
the NT state and the empty metallofullerence state in the gap of the
nanotube. 

%
%

\subsubsection{On-site energy (asymmetry between two
   sublattices)}\label{subsec:em-di}

We now briefly comment on the on-site interaction.
The on-site interaction is defined by
\begin{equation}
 {\cal H}_{\rm site} = \sum_{i} \epsilon(r_i) a_i^\dagger a_i,
\end{equation}
where $\epsilon(r_i)$ denotes the on-site energy.
The constant component of the on-site energy determines the origin of
the energy and does not affect the dynamics.
By contrast, its modulation part can impart physical effects on the 
conducting electrons.
The modulation can be divided into a symmetric part and an asymmetric
part for the two sublattices.
They appear in the low-energy Hamiltonian as the coefficients of
$\sigma_0$ (the identity matrix) and $\sigma_3$~\cite{VAA},
respectively, in the form
\begin{equation}
 {\cal H}_{\rm K} \to
  v_F \sigma \cdot (p-A) + G(r) \sigma_0 + P(r)\sigma_3.
\end{equation}
The former, $G(r)$, acts as the local electrical potential and the
latter, $P(r)$, represents an asymmetric potential between the two
sublattices.
While $G(r)$ does not change the energy {\it gap} of the theory, $P(r)$
does. 
The local energy gap along the axis can be calculated using dimensional
reduction as
\begin{equation}
 E_{\rm gap}(x_2) = 2 \sqrt{ v_F^2 
  \left( p_1^0 - \oint \frac{dx_1}{|C_h|}A_1 \right)^2 
  + \left( \oint \frac{dx_1}{|C_h|} P \right)^2 }.
\end{equation}
We note that the deformation-induced gauge field can increase or
decrease the energy gap, depending on the sign of $A_1$.
However, the asymmetric potential always increases the energy gap.

\section{Mixing of Fermi points}\label{sec:mixing}

\begin{figure}[htbp]
 \begin{center}
  \psfrag{kx}{$k_x$}
  \psfrag{ky}{$k_y$}
  \psfrag{p1}{$p_1$}
  \psfrag{+A}{$+A_1^{\rm em}$}
  \psfrag{-A}{$-A_1^{\rm em}$}
  \psfrag{kF}{$+k_F$}
  \psfrag{-kF}{$-k_F$}
  \includegraphics[scale=0.5]{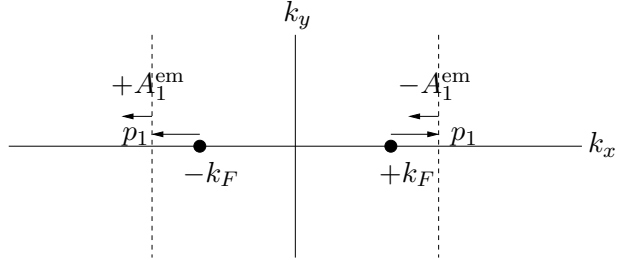}
 \end{center}
 \caption{Momentum representation used in \S~\ref{sec:mixing}. We
 consider zigzag nanotubes, whose Fermi points denoted as solid circles,
 are located on the $k_x$ axis.
 The dashed lines represent the two energy bands near the K and K'
 points.} 
 \label{fig:mixing}
\end{figure}

To this point, we have disregarded the matrix element that causes mixing
of the two Fermi points K and K', whose effect is on the order of
$\delta V_a (2k_F)$.
In this section, we set the Fermi velocity, $v_F$, to unity.
We begin by introducing the Schr\"odinger equation
\begin{align}
 i \hbar \frac{\partial}{\partial t}
 \begin{pmatrix} \psi^{\rm K} \cr \psi^{\rm K'} \end{pmatrix} =
 \begin{pmatrix} {\cal H}_{\rm K} & 0 \cr 0 & ({\cal H}_{\rm K'})' 
 \end{pmatrix}
 \begin{pmatrix} \psi^{\rm K} \cr \psi^{\rm K'} \end{pmatrix},
\end{align}
where we have defined $({\cal H}_{\rm K'})'$ as ${\cal H}_{\rm K'}$
[see Eq.~(\ref{eq:eff-dynamics})] with the replacement 
$(p_1,p_2) \to (-p_1,p_2)$.
Thus, each diagonal block is given, respectively, by
${\cal H}_{\rm K} = \sigma_1 (p_1-A_1-A_1^{\rm em}) 
+ \sigma_2 (p_2- A_2-A_2^{\rm em})$ and 
$({\cal H}_{\rm K'})' = \sigma_1 (p_1 - A_1 + A_1^{\rm em}) 
+ \sigma_2 (p_2 +A_2-A_2^{\rm em})$.
Roughly speaking, $p_1$ measures the difference between the momenta
(about the axis) for the K and K' states (Fig.~\ref{fig:mixing}).
The $\sigma_1$ terms serve as mass terms (or the energy gap) when we
consider an effective one-dimensional model.
In this regard, we note the different signs for an external magnetic
field $A_1^{\rm em}$. 
Because $A_1^{\rm em}$ can be tuned by a magnetic field {\it along} the 
axis, it can cause an asymmetry between the masses of the two
modes~\cite{AA}; that is, it can break the time-reversal symmetry.

The matrix defined above acts on the four component wave function
\begin{align}
 \begin{pmatrix} \psi^{\rm K} \cr \psi^{\rm K'} \end{pmatrix}
 = \begin{pmatrix} \psi^{\rm K}_A \cr \psi^{\rm K}_B \cr 
   \psi^{\rm K'}_A \cr \psi^{\rm K'}_B \end{pmatrix} = \psi,
\end{align}
where the two components of $\psi_{\rm K}$, $\psi_A^{\rm K}$ and
$\psi_B^{\rm K}$, and of $\psi_{\rm K'}$, $\psi_A^{\rm K'}$ and
$\psi_B^{\rm K'}$, correspond to the wave functions at the two
sublattices, A and B.
The $\psi^{\rm K}_A$ component couples to the wave functions on the
B-sublattice, $\psi^{\rm K}_B$ and $\psi^{\rm K'}_B$, and the coupling
between $\psi^{\rm K}_A$ and $\psi^{\rm K'}_B$ creates the mixing 
effect. 
Therefore, the general dynamics (except for the topological defects
mentioned below) can be described by the Hamiltonian
\begin{align}
 {\cal H}_{\rm eff} = 
 \begin{pmatrix} 0 & ({\cal H}_{\rm K})_{12} & 0 & 2b_1 \cr
  ({\cal H}_{\rm K})_{21} & 0 & 2b_2 & 0 \cr  
  0 & 2b_2^* & 0 & ({\cal H}_{\rm K'})'_{12} \cr
 2b_1^* & 0 & ({\cal H}_{\rm K'})'_{21} & 0 \end{pmatrix},
 \label{eq:ham-defo}
\end{align}
where $b=(b_1,b_2)$ are complex functions.

The mixing of Fermi points may have a strong effect on the dynamics,
especially when there exist pentagons or heptagons in the
surface~\cite{GGV}.
We attempt to construct an effective low-energy model including the effects
of a topological defect and of a surface deformation.
For this purpose, we first note that the effective dynamics of
Eq.~(\ref{eq:ham-defo}) can be expressed as a special case of the
Schr\"odinger equation
\begin{align}
 \sigma^\mu D_\mu \psi = 0,
 \label{eq:g-weyl}
\end{align}
where $D_\mu \equiv (p_\mu - A_\mu^0) \tau_0 - A_\mu^i \tau_i$ is the
covariant momentum and the summation variable $\mu$ is understood to
take the values 0, 1 and 2.
Here, $A_\mu^i$ is a generalization of the previously defined
deformation-induced gauge field, and $\tau_i$ represents the Pauli
matrices (with $\tau_0$ the identity element) acting on the wave
functions ${}^t(\psi^{\rm K},\psi^{\rm K'})$.
We have defined $\sigma^\mu = \sigma_\mu$ and $p_0 = i\hbar \partial_t$,
where $t$ is the time variable.
The Schr\"odinger equation is formally equivalent to the Weyl equation
in {\it U}(1) Abelian and {\it SU}(2) non-Abelian deformation-induced
gauge fields.

The Hamiltonian of Eq.~(\ref{eq:ham-defo}) can be obtained by setting
the generalized gauge field of Eq.~(\ref{eq:g-weyl}) as
\begin{align}
 & A_1^3 = A_1^{\rm em}, \ A_2^0 = A_2^{\rm em}, \
 A_1^0 = A_1, \ A_2^3 = A_2, \nn \\
 & A_1^1 = - \Re (b_1+b_2), \ A_1^2 = \Im (b_1+b_2), \
 A_2^2 = \Re (b_1-b_2), \ A_2^1 = \Im (b_1- b_2),
\end{align}
and setting other components to zero.
Concerning the relationship between a pentagon or heptagon and the
non-Abelian deformation-induced gauge field, the main contribution from
a pentagon or a heptagon is the interchange of the wave functions of the
two Fermi points.
This occurs when a low-energy electron moves around topological
defect~\cite{GGV}, which can be expressed in terms of a nonvanishing
time component of the matrix element $\tau_1$ or $\tau_2$ part as
\begin{equation}
 \begin{pmatrix} 0 & \sigma^0 (A_0^1 -i A_0^2) \cr 
 \sigma^0 ( A_0^1 + i A_0^2) & 0 \end{pmatrix}
 \begin{pmatrix} \psi^{\rm K} \cr \psi^{\rm K'}\end{pmatrix}.
\end{equation}
This is because a pentagon (heptagon) creates a coupling between
$\psi^{\rm K}_A$ and $\psi^{\rm K'}_A$, and between $\psi^{\rm K}_B$ and
$\psi^{\rm K'}_B$. 
Hence, the local dynamics of the low-energy conducting electron on a
deformed (graphite) surface is governed by Eq.~(\ref{eq:g-weyl}), and
the lattice structure fixes the Abelian and non-Abelian
deformation-induced gauge field. 


\section{Discussion}\label{sec:discussion}

The main result for the local energy gap given in
Eq.~(\ref{eq:band-gap}) is determined by only the configuration of the
deformation-induced gauge field. 
This is due to the special energy dispersion relation of the graphite
sheet, which allows us to predict several important physical
consequences, not by solving the Hamiltonian explicitly but from the
gauge field configuration itself.
To understand the configuration of the deformation-induced gauge field,
we have classified the geometries into two types: a bond-bending
deformation and a bond-stretching deformation.
For the bond-bending type, we can extract the gauge field from the
geometric information provided by the normal vector as we did in
\S~\ref{subsec:b-class}, where no microscopic theory is necessary 
to calculate the field configuration.
By contrast, for the bond-stretching type, we need a microscopic
theory capable of predicting the bond length or the gauge field in order
to calculate the local energy gap (\S~\ref{subsec:s-class}).
We can carry out a qualitative analysis of the bond-stretching type
using a microscopic model and compare it with our local energy gap
formula. 
Such a study will be reported elsewhere.

When we consider the mixing effect due to a short-range interaction, we 
should generalize the geometry-induced gauge field from an Abelian to a 
non-Abelian field. 
Constructing a general low-energy theory is of prime importance, because
a general deformation may generate a topological defect, such as a
pentagon or heptagon, which can mix the wave functions at the two Fermi
points.
In addition, we can consider a higher genus material whose kinematics
are nontrivial~\cite{SKS,Terrones}.
One possible way to examine such a material is to extract useful
information from a dynamical model.

Finally, we comment on the wave functions in nanotubes with a local
energy gap modulation.
First, note that we did not solve the effective Hamiltonian of
Eq.~(\ref{eq:eff-dynamics}), but, instead, we extracted the energy gap
along the axis by considering the spectrum of the effective
one-dimensional theory, given by Eq.~(\ref{eq:effective-h}), obtained by
dimensional reduction.  
Solving the equation for a general $A$ without dimensional reduction
would be difficult, even in the absence of mixing, because the
eigenfunctions of the Hamiltonian, ${\cal H}_{\rm K} \psi = E \psi$,
should also satisfy~\cite{Jackiw} the equation
\begin{align}
 {\cal H}^2_{\rm K} \psi 
 = \left[ 
 \left( p - A \right)^2 \sigma_0 + \hbar B_\perp \sigma_3 \right] \psi 
 = E^2 \psi,
\end{align}
where we have set $v_F =1$ and $A^{\rm em} = 0$.
The spectrum of the first term on the right-hand side is known to be
quite nontrivial, even for simple vortex configurations~\cite{AB}. 
Furthermore, it is already difficult to solve the one-dimensional
Hamiltonian, due to the local modulation mass term. 
The nature of the wave function should be important for investigating
the possibility of ``energy gap engineering''~\cite{Lee}, which requires
multiple quantum dots prepared by a local modulation of the energy gap.

\section{Conclusion}\label{sec:conclusion}

We have examined the effects of surface deformation on the ground state
of conducting electrons using the nearest-neighbor tight-binding
Hamiltonian. 
Within the framework of the effective-mass theory, we clarified the
relationship between a local deformation of the lattice and the local
energy gap along the axis in terms of the deformation-induced magnetic 
field.
We formulated an effective theory describing the dynamics on a general
deformed surface, including topological defects that can cause a mixing
of the two Fermi points. 
The theory is formally equivalent to the Weyl equation in
{\it U}(1) Abelian and {\it SU}(2) non-Abelian deformation-induced gauge
fields.

\section*{Acknowledgements}
K. S. acknowledges support from the 21st Century COE Program of
the International Center of Research and Education for Materials at
Tohoku University.
R. S. acknowledges Grants-in-Aid (Nos. 13440091 and 16076201) from the
Ministry of Education, Culture, Sports, Science and Technology, Japan.


\end{document}